\documentclass[letterpaper,english,aps, prl, reprint, superscriptaddress]{revtex4-2}
\usepackage[T1]{fontenc}
\usepackage[latin9]{inputenc}
\usepackage{verbatim}
\usepackage{amsmath}
\usepackage{amssymb}
\usepackage{graphicx}
\usepackage{relsize}
\usepackage{esint}
\usepackage{multirow}
\usepackage{adjustbox}
\usepackage{mathrsfs}
\usepackage{xcolor}

\usepackage{hyperref}
\hypersetup{
	colorlinks = true,
	allcolors = blue
}

\usepackage{times}

\makeatletter
\pdfpageheight\paperheight
\pdfpagewidth\paperwidth

\PassOptionsToPackage{position=top}{subfig}

\makeatother

\let\vec\boldsymbol
\usepackage{babel}
\usepackage{mathrsfs}

\definecolor{darkgreen}{rgb}{0,0.5,0}
\begin{document}

\title{Anomalous universal quantum transport in 2D asymptotic quasiperiodic system}
\author{Ting-Fung Jeffrey Poon}
\affiliation{International Center for Quantum Materials, School of Physics, Peking University, Beijing 100871, China}
 \affiliation{Hefei National Laboratory, Hefei 230088, China}
\author{Yuhao Wan}
\affiliation{International Center for Quantum Materials, School of Physics, Peking University, Beijing 100871, China}
\author{Yucheng Wang}
\thanks{Corresponding author: wangyc3@sustech.edu.cn}
\affiliation{Shenzhen Institute for Quantum Science and Engineering,
Southern University of Science and Technology, Shenzhen 518055, China}
\affiliation{International Quantum Academy, Shenzhen 518048, China}
\affiliation{Guangdong Provincial Key Laboratory of Quantum Science and Engineering, Southern University of Science and Technology, Shenzhen 518055, China}
\author{Xiong-Jun Liu}
\thanks{Corresponding author: xiongjunliu@pku.edu.cn}
\affiliation{International Center for Quantum Materials, School of Physics, Peking University, Beijing 100871, China}
%\affiliation{Collaborative Innovation Center of Quantum Matter, Beijing 100871, China}
%\affiliation{Beijing Academy of Quantum Information Science, Xibeiwang East Rd, Beijing 100193, China}
%\affiliation{CAS Center for Excellence in Topological Quantum Computation, University of Chinese Academy of Sciences, Beijing 100190, China}
\affiliation{Hefei National Laboratory, Hefei 230088, China}
\affiliation{International Quantum Academy, Shenzhen 518048, China}

\begin{abstract}
Quasiperiodic systems extend the concept of the Anderson transition to quasi-random and low-dimensional realms and have garnered widespread attention. Here, we propose the asymptotic quasiperiodic two-dimensional systems characterized by a sequence of %periodic lattice models with a quasiperiodic limit and explore a series of
rational magnetic fluxes, which have an irrational limit, and predict exotic universal wave-packet dynamics and transport phenomena associated with the asymptotic quasiperiodicity (AQP). The predictions unveil a class of multiple metal-insulator transitions driven by a novel interplay effect between AQP, relaxation, and finite temperature, which further reveals a unified and profound mechanism. Specifically, all the transport phenomena, including the wave-packet dynamics, the bulk and edge transport, are unified in the universal scaling laws unveiled in the asymptotic quasiperiodic regime, which demonstrate a nontrivial asymptotic connection to quantum phases in the quasiperiodic limit. %reveal the impact of asymptotic quasiperiodicity (AQP) together with relaxation on transport phenomena.
%Specifically, we demonstrate anomalous bulk transport with universal scaling characteristics in the wave-packet dynamics and conductivity, and predict novel interplay effects involving AQP, temperature, and relaxation, leading to unprecedented multiple anisotropic metal-insulator transitions. The AQP also leads to the anisotropic edge tunneling transport.
Our work enriches the universal quantum transport phenomena, adds to the basic mechanisms underlying metal-insulator transitions, and opens up an avenue to study the exotic transport physics with AQP in high dimensions. %, providing new perspectives and approaches for the study of two-dimensional transport physics}.
\end{abstract}
\maketitle
%14,22; 1,3,8,9,13,16,18,23; 7,20
\textcolor{blue}{\em Introduction.}--Quasiperiodic systems~\cite{AL4,AL5}, positioned between periodic and disordered structures, have garnered widespread attention in recent years, particularly in the context of lower-dimensional Anderson localization~\cite{AL5, ALLD6, ME2, ME4, ME5, ME6, ME7, ME8, ME9, BG7, ME0, MEa, MEb, MEc, MEd,MFS3, MFS0, MFSa, MFSb, MFSc, MFSd, MFSe, MFSf, MFSg, MFSh, MFSi}, mobility edges~\cite{ME2, ME4, ME5, ME6, ME7, ME8, ME9, BG7, ME0, MEa, MEb, MEc, MEd}, multifractal critical phases~\cite{MFS3, MFS0, MFSa, MFSb, MFSc, MFSd, MFSe, MFSf, MFSg, MFSh, MFSi}, Bose glasses~\cite{BG, BG2, BG3, BG4, BG5, BG6, BG8, BG9, BG0}, and many-body localization~\cite{MBL, MBL2, MBL3, MBL4, MBL7, MBL9, MBLa, MBLc}. Accordingly, quasiperiodic systems show diverse transport phenomena even in one dimension, including localized, subdiffusive, diffusive, superdiffusive, and ballistic transport~\cite{Tp, Tp2, Tp3, Tp4}. This contrasts sharply with one-dimensional (1D) disordered systems, which are always localized regardless of disorder strength~\cite{1DDisorder, 1DDisorder2, 1DDisorder3}. Similarly, in 2D systems, it is well-established that bulk wave functions become localized with weak disorder~\cite{AL2,AL3}. This observation, along with the distinct transport physics in 1D disordered and quasiperiodic systems, prompts one to consider 2D quasiperiodic systems~\cite{2DQS, 2DQS2, 2DQS3, 2DQS4, Moire1, Moire5, Moire6} to explore unique transport phenomena different from both periodic and random systems. 

A most quintessential 2D phenomenon is the integer quantum Hall effect observed in electronic gases subjected to strong perpendicular magnetic fields, where the Hall conductivity is quantized to integer (Chern number) multiples of $e^2/h$, with vanishing longitudinal conductivity, revealing its topological properties~\cite{QHE}. The 2D quasiperiodic Hall systems are obtained when the magnetic flux $2\pi\phi$ per unit cell is irrational. In additional to  synthetic quantum systems like optical lattices, in which the synthetic magnetic fluxes can be generated through light-atom couplings, recently the 2D twisted Moir\'{e} materials show also feasibility in tuning the flux $\phi$ due to much enlarged unit cells and attracted considerable interests~\cite{Moire2, Moire3, Moire4, Moire7,DEParker}. In the real experiment, an exact irrational flux is hard to engineer. For a rational magnetic flux $\phi=p/q$, the system is effectively quasiperiodic when the magnetic unit cell, determined by $q$, is larger than system size, but it is always periodic in the thermodynamic limit. %The system with rational flux in thermodynamic limit is generally periodic, and
An intriguing open issue is that, whether the transport physics in 2D system with irrational magnetic flux can be studied and probed from that in  systems with rational flux in the thermodynamic limit? %It is therefore intriguing to explore whether universal transport physics can be obtained by tuning the flux to be asymptotically incommensurate, as readily accessible in 2D twisted materials.

In this Letter, we propose the 2D asymptotic quasiperiodic Hall systems, which are characterized by a rational sequence of magnetic fluxes $\phi_n=p_n/q_n$, with only the limit $\phi_\infty$ being irrational, and predict the interplay effects of the asymptotic quasiperiodicity (AQP), relaxation, and temperature, which lead to unprecedented multiple anisotropic metal insulator transitions (MITs). The qualitatively nontrivial transport phenomena with rational $\phi_n$, including wave packet dynamics, bulk and edge transport, are observed and have a unified profound mechanism interpreted as universal scaling laws, which are deeply connected to the quantum phases in the quasiperiodic limit characterized by $\phi_\infty$. %as a consequence of the competition of the AQP and relaxation. %exhibit universal scaling behavior, leading to the , which originate from the competition of relaxation and asymptotic quasiperiodicity,
The predicted new multiple MITs are different from those driven by the celebrated mechanisms~\cite{metal8, metal9}, including by spontaneous symmetry breaking~\cite{ssb, ssb2, ssb3, ssb4}, Lifshitz transition~\cite{metal0, metalb}, Mott gap in Hubbard models~\cite{Mott}, and by the (quasi-)disorders~\cite{AL2, rqr, rqr2}. These results highlight the exotic universal transport physics in the asymptotically quasiperiodic 2D systems, and can be applied to the recently attractive 2D Moir\'{e} materials.

\begin{figure}[!ht]
\centerline{\includegraphics{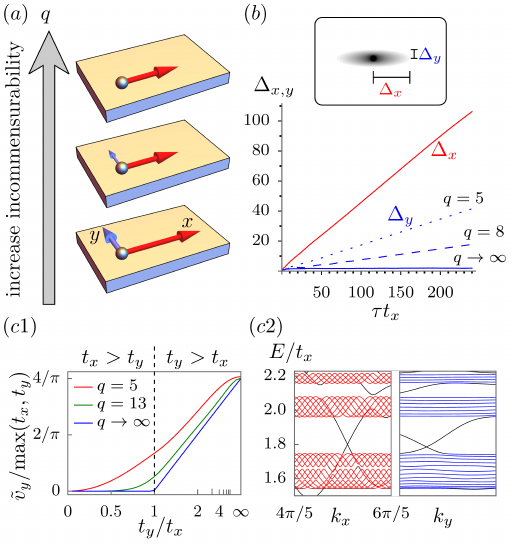}}
\caption{(a) Schematic diagram showing how AQP impacts transport behavior for $t_x>t_y$. As $q$ increases, the system becomes more localized (denoting by shorter arrowed lines) in the $y$-direction, while remaining at a nonzero value along $x$ even as $q\rightarrow\infty$. (b) $\Delta_{x,y}$ vs time. $\phi = 3/5, 5/8$ and $(\sqrt{5}-1)/2$ for the dotted, dashed and solid blue lines, respectively. (c1) $\tilde{v}_y$ vs hopping. $\phi = 3/5 (8/13)$ for the red (green) line. The blue line represents both $\phi=144/233$ and $q\rightarrow\infty$ that overlaps. (c2) The zoomed-in band structure vs $k_x$ (red) and $k_y$ (blue) for $\phi=(\sqrt{5}-1)/2$. $t_x=1$ and $t_y=0.7$ for (b,c2). The black lines represent the edge states at the gap.} \label{fig: BulkT}
\end{figure}

\textcolor{blue}{\em Model.}--We start with a 2D square lattice model with an asymptotically incommensurate magnetic flux $2\pi\phi$ threading each plaquette. By taking the gauge $\vec A=A_d\hat e_d$ with $A_d=2\pi\phi j_{\bar d}$, the Hamiltonian can be written as
%can demonstrate the exotic transport behaviors
\begin{equation}
	\label{modelH}
	H = \sum_{j_dj_{\bar{d}}} t_d e^{-2\pi i\phi j_{\bar{d}}} c_{j_d+1,j_{\bar{d}}}^\dagger c_{j_dj_{\bar{d}}} + t_{\bar{d}} c_{j_d,j_{\bar{d}}+1}^\dagger c_{j_dj_{\bar{d}}} + h.c.,
\end{equation}
where $d$ represents the $x$ or $y$ direction, $\bar{d} \perp d$, $t_d$ is the hopping along $d$-direction, and $\phi = \phi_n = p_n/q_n$ is rational, except for the finite irrational limit $\phi_\infty$. For example, $\phi_\infty=(\sqrt{5}-1)/2$ can be approximated via Fibonacci sequences~\cite{GR}. After performing Fourier transformation along $d$-direction, we have $H = \sum_{k_d}H_{1D}(k_d)$, where $H_{1D}(k_d)=\sum_{j_{\bar{d}}} 2t_d\cos(2\pi\phi j_{\bar{d}}+k_d)  c_{k_dj_{\bar{d}}}^\dagger c_{k_dj_{\bar{d}}} + t_{\bar{d}} \left(c_{k_d,j_{\bar{d}}+1}^\dagger c_{k_dj_{\bar{d}}}\right)+h.c.$ gives the Aubry-Andr\'{e}-Harper model for irrational $\phi = \phi_\infty$ ~\cite{AL5}. Then, for $t_d>t_{\bar{d}}$ the system is localized (extended) along the $\bar{d}$-direction ($d$-direction)~\cite{AAnote}. %and there is a critical point $t_d=t_{\bar{d}}$ at which both directions are critical~\cite{SM}.
Note that if we rewrite Eq.~\eqref{modelH} in the $A_{\bar d}$ gauge, the system is translational invariant in $\bar d$ direction. As gauge transformation does not change physics, this leads to a puzzle: why the states are localized in the translational invariant $\bar d$-direction? We shall resolve this puzzle below by revealing a novel mechanism for Anderson localization in this 2D system. More importantly, instead of focusing on the quasiperiodic limit $\phi=\phi_\infty$, here we consider AQP regime with fluxes varying according to the rational sequence $\phi_n=p_n/q_n$ [Fig. \ref{fig: BulkT}(a)], and explore the new universal transport physics regarding the sequence, beyond the extended and localization phases. %, which deeply connects to the quantum phases in $x$ and $y$ directions in the quasiperiodic limit.}
%We below address a more important issue: As $\phi$ gradually varies from rational to irrational numbers, are
%there any unprecedented universal characteristic pattern for the transport physics in AQP regime beyond the expected transition from the extended to the localized phase in conventional quasiperiodic systems [Fig. \ref{fig: BulkT}(a)]? Also, whether the relaxation and finite temperature can bring nontrivial and universal effects on the transport in such regime?

\textcolor{blue}{\em Bulk transport.}--We first consider the coherent wavepacket dynamics. Let the wavepacket $\psi(0)$ be initially centered at $(x_0,y_0)$, and we examine the directional mean-square displacement at time $\tau$ defined by $\Delta_d(\tau) = \sqrt{\langle (d-d_0)^2\rangle_{\psi(\tau)}}$ for the wavefunction $\psi(\tau)=e^{-iH_y\tau}\psi(0)$. Then for $t_x>t_y$ we have the generic relation ($q=q_n$)
\begin{equation}\label{bulk1}
	\Delta_x = C_1, \tau; \ \Delta_y = \begin{cases}
		C_2 \tau, & \text{for rational }\phi=\frac{p}{q} \\
		\widetilde{C}_3, & \text{for irrational }\phi
	\end{cases}
\end{equation}
where the constants $C_{1,2}$ depends on $q$. We set $\psi(0) = \mathcal{N} e^{-((j_x-x_0)^2+(j_y-y_0)^2)/\Delta_0}$, with $\Delta_0$ being initial wave-packet width, and simulate the average over $x_0$ and $y_0$, and find that the coefficients exhibit distinct universal scalings
	\begin{equation}\label{bulk2}
		\langle C_2\rangle \sim \widetilde{C}_2 t_x(t_y/t_x)^{q},\   \langle C_1\rangle \sim \widetilde{C}_1 (t_x-t_y) + q^{-1}\widetilde{C}_4\sqrt{t_xt_y}
	\end{equation}
with $\tilde{C}_{1,2,3,4}$ being of order $1$~\cite{SM} and $\langle\cdot\rangle $ denoting the average of $p$ and $\psi(0)$. The displacement speed $\langle C_2 \rangle$ exhibits a clear exponential law versus $q$. This is because the ballistic transport along $y$ direction can be reached only after every $q$-site hopping, corresponding to the $q$-th order perturbation process. %proportional to the effective hopping over $q$ sites, decays exponentially with $q$, as periodicity only occurs after a distance of $q$ sites, within which the system is localized.
In contrast, $\langle C_1 \rangle$ consists of two terms, one independent of $q$ and the other having a power-law relationship with $q$ (details in supplementary material~\cite{SM}). The distinct scaling laws in the ballistic transport in two directions for finite $q$ show nontrivial connection to the localization (extended) phase in the $y(x)$-direction in the incommensurate limit $q\rightarrow\infty$ [Figs. \ref{fig: BulkT}(a,b)]. We then examine the maximal group velocity in the $y$-direction and averaged over all bands. %and $p$ characterizes the particles' maximal transport power.
After some algebra we can obtain that [Fig. \ref{fig: BulkT}(c1)]~\cite{SM}
%\begin{equation}\label{bulk3}
%	\tilde{v}_y^{(\infty)} \defeq \tilde{v}_{y} (q\rightarrow\infty) = \begin{cases}
%		0 & \text{for }t_x>t_y, \\
%		\frac{4}{\pi} (t_y-t_x) & \text{for }t_x<t_y.
%	\end{cases}
%\end{equation}
%$\beta \gtrsim 1.77$
%Numerical analysis for finite $q$~\cite{SM} shows that
\begin{equation}\label{bulk4}
	\tilde{v}_y \sim  \begin{cases}
		t_x(t_y/t_x)^{q/2}, & \text{for }t_x>t_y, \\
		\frac{4}{\pi} (t_y-t_x)+\tilde C_5q^{-1} \sqrt{t_x t_y}, & \text{for }t_x<t_y,
	\end{cases}
\end{equation}
with $\tilde C_5$ being of order $1$. The scaling exponent for $t_x>t_y$ is $q/2$ rather than $q$. This %is because the particles with maximal ability of transport ``see" an effective periodicity to be $q/2$.
is understood in the following way. We transform the Hamiltonian in $A_y$ gauge to $k$-space in both directions, yielding $H= \sum_{k_x,k_y} 2t_x\cos (k_x) c^\dagger_{k_xk_y} c_{k_xk_y} + (t_ye^{ik_y} c^\dagger_{k_x+2\pi\phi,k_y} c_{k_x,k_y} + h.c.)$. Then $\tilde{v}_y$ is proportional to maximum bandwidth of subbands versus $k_y$, parameterized by $k_x$. The original band splits into $q$ subbands due to the transitions between states with $k_x$ and $k_x+2\pi\phi$. For this the $q$ states with diagonal potentials $2t_x\cos\tilde k_x$ ($\tilde k_x=k_x, k_x+2\pi\phi,..., k_x+2\pi(q-1)\phi$) are coupled together. The bandwidths determined by the maximum transitions are obtained from the set of $q$ states with $k_x=0$, which separate into two equal groups, with each including $q/2$ distinct potentials. Thus the $q/2$-th order perturbation [$\sim(t_y/t_x)^{q/2}$] determines the bandwidth and maximal group velocity~\cite{SM}.

 With Eqs. (\ref{bulk1}-\ref{bulk4}) we reach a new mechanism for the Anderson localization in $y$ direction, which is translational invariant under $A_y$ gauge, in the case of $t_x>t_y$ and $\phi=\phi_\infty$. %through the $A_y$ gauge condition with $k_y$ being a good quantum number. %However, it is evident that we can choose the $A_y$ gauge such that $k_y$ is a good quantum number. To reconcile both facts,
In Fig.~\ref{fig: BulkT}(c2) we plot the energy dispersion versus $k_x$ and $k_y$, in $A_x$ and $A_y$ gauges, respectively. In contrast to the dispersive bands with $k_x$, all the bulk subbands become flat versus $k_y$, consistent with the fact that $\tilde{v}_y$ exponentially decays to zero with increasing $q$ and $\tilde{v}_y(q\rightarrow\infty)=0$. This shows a novel mechanism that Anderson localization in the Aubry-Andr\'{e} model (in $A_x$ gauge) is equivalently mapped to 1D Bloch flat bands (in $A_y$ gauge) through gauge transformation.
	
\begin{figure}[!ht]
\centerline{\includegraphics{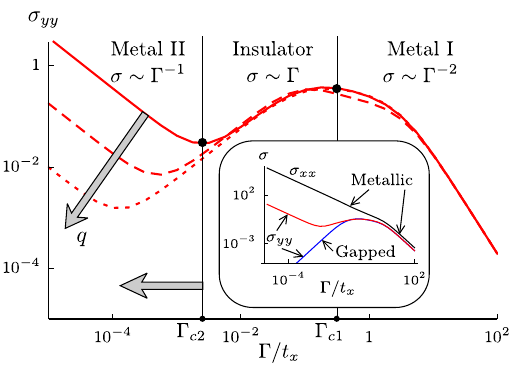}}
\caption{$\sigma_{yy}$ vs $\Gamma$, where $t_x=1, t_y=0.7$. $\phi=8/13, 11/17$, and $13/21$ for the solid, dashed and dotted red lines, respectively.  The two gray arrows in the Metal II region indicate that as $q$ increases, $\Gamma_{c2}$ approaches $0$ exponentially. Inset: Black: $\sigma_{xx}$. Red (Blue): Typical $\sigma_{yy}$ if $E_F$ is at the band (inside gap for QHE regime).}
\label{fig: 0Tsigma}
\end{figure}

We then investigate the universal bulk transport phenomena in the presence of relaxation at zero temperature. The DC longitudinal conductivity with relaxation is written as 
\begin{equation}\label{kubolongit}
\sigma_{dd} = \frac{e^2}{h}\int \frac{d^2k}{(2\pi)^2} \mathfrak{R} \left\{\text{tr} \left[\left(G_0^+-G_0^-\right) \frac{d\mathcal{H}}{dk_d} G_0^- \frac{d\mathcal{H}}{dk_d}\right]\right\},
\end{equation}
where $\mathfrak{R}\{\cdot\}$ denotes the real part, $G_0^\pm = (E_F\pm i\Gamma-\mathcal{H})^{-1}$, $\mathcal{H}$ is Fourier transformation of $H$ into 2D $k$-space, and $\Gamma$ denotes a finite relaxation rate, which originates from a uniform spatial decay effect~\cite{KuboBand, KuboBand2, KuboBand3,SM}, such as the coupling of a 2D material to a substrate~\cite{scattering1,scattering2} or the introduction of losses at each site in cold atom systems~\cite{scattering3,scattering4}. Examination on $\sigma_{yy}$ reveals that there are three qualitatively different regions [Fig.~\ref{fig: 0Tsigma}]~\cite{SM}: (A) Metal I. $\sigma_{yy} \propto \Gamma^{-2}$ in the region with sufficiently large $\Gamma > \Gamma_{c1}$; (B) Metal II. $\sigma_{yy} \propto \Gamma^{-1}$ in the region with sufficiently small $\Gamma < \Gamma_{c2}$; (C) Insulator. $\sigma_{yy} \propto \Gamma^{\alpha}$, where $\alpha \approx 1$ at the intermediate region. The scalings of $\sigma_{yy}$ in regions A and B can be proven by asymptotic behaviors of Eq.~\eqref{kubolongit}. For large $\Gamma$, $\sigma_{yy} = e^2(2h\Gamma^2\pi^2)^{-1}\int dk_xdk_y \text{ tr } [\mathcal{H}^2]$; for small $\Gamma$, $\sigma_{yy} = e^2(4h\Gamma\pi^2)^{-1} \int dk_xdk_y \sum_b \delta(\epsilon_b(k_x,k_y)-\mu)(\partial\epsilon_b/\partial k_\alpha)^2$, where $\epsilon_b$ is the energy of the $b-$th band. The critical relaxation rates $\Gamma_{c1,c2}$ satisfy the scalings~\cite{SM}
\begin{equation}\label{scaling}
\Gamma_{c1} \sim t_y; \ \Gamma_{c2} \sim t_x(t_y/t_x)^q,
\end{equation}
which are universal and have profound mechanism. Note that $1/\Gamma$ denotes a time scale, during which each particle is scattered once. The flux $\phi$ splits the original Bloch band into $q$ magnetic bands, which dominate the ballistic bulk transport in the weak relaxation limit, giving a band metal phase (Metal II),  in which $\sigma_{yy}\propto\Gamma^{-1}$, similar to Drude model. In the opposite strong $\Gamma$ limit, the phase information associated with magnetic flux is wiped out, so the band splitting between the magnetic bands is removed. Then the ballistic transport is dominated by the original Bloch band results, rendering the Metal I region. 
In this region, the conductivity is $\sigma_{yy}\propto\Gamma^{-2}$, which arises from both the dephasing and decay of particles in the case of strong relaxation (each effect contributes to a factor $\Gamma^{-1}$ of $\sigma_{yy}$). The most nontrivial region is in-between, in which the relaxation and AQP effect compete, rendering the mechanism of the universal scalings at the critical transitions. The particles see the magnetic unit cell only after it can coherently hop $q$ sites without being fully scattered since the magnetic unit cell is expanded by $q$ times. This leads to a critical relaxation rate $\Gamma_{c2}$ proportional to $q$-th order of the hopping couplings, i.e. $\sim (t_y/t_x)^q$, beyond which the coherent hopping of particles is limited within a unit cell, and then they experience an effective quasiperiodic system, instead of periodic system with magnetic bands, giving the intermediate insulating phase.  We see that $\Gamma_{c2}$ decays exponentially with $q$, so a tiny $\Gamma$ drives the periodic Hall system to be effectively quasiperiodic at relatively large $q$. In the incommensurate limit $q\rightarrow\infty$, the Metal-II phase disappears and $\sigma_{yy}$ vanishes for $\Gamma\rightarrow0$, but is finite for $\Gamma>0$.
Furthermore, if the particles are fully scattered within every single hopping process (with time scale $1/t_y$), the effective quasiperiodicity is further removed, giving the other critical point $\Gamma_{c1}\sim t_y$. In the $\Gamma>\Gamma_{c1}$ regime, the bulk reenters metallic phase generically, even $E_F$ is in the band gap [insert of Fig.~\ref{fig: 0Tsigma}]. Finally, in the Supplementary Material~\cite{SM}, we also show that the insulating region turns to a critical metal phase with $\sigma_{yy}=\sigma_{xx}\sim\Gamma^{-1/2}$ in the isotropic regime $t_x=t_y$. To observe these results in real experiment requires that the system size be large compared to the size ($q$) of a unit cell.

%Moreover, the bulk transport at $q\rightarrow\infty$ also behaves differently from conventional QHE. In the absence of relaxation, system with QHE has zero longitudinal conductivity in both direction, and the introduction of the lattice length scale allows the longitudinal conductivity (i.e. Hofstadter's butterfly). However, the presence of irrational $\phi$ forces one of the longitudinal conductivities to be zero, albeit the presence of lattice. Likewise, in the presence of disorder, away from the quantized plateau of transverse conductivity, $\sigma_{xx}$ ($\sigma_{yy}$) decreases (increases) on increasing disorder, while conventional QHE predicts both to be increasing.

\begin{figure}[!ht]
\centerline{\includegraphics{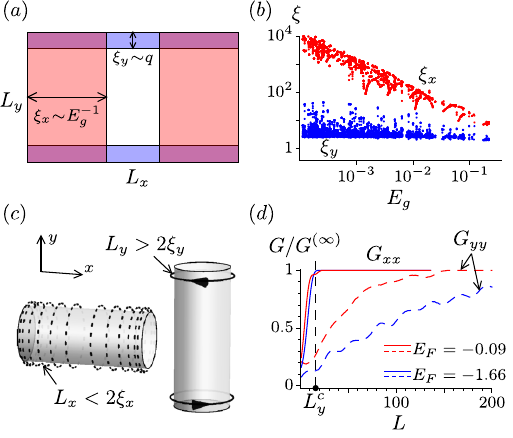}}
\caption{(a) Schematic diagram showing $\xi_d$ of the edge states. (b) $\xi_{x,y}$ vs $E_g$. Energy of the edge states are set at gap center. Data includes $\phi=p/q$ with coprime $p$ and $q$, $5\leq q\leq 149$, and $p$ minimizing $|\phi - (\sqrt{5}-1)/2|$. (c) Schematic diagram shows the effect of anisotropy of the edge states. Edge states in $x$-periodic system (right cylinder) is localized on the edge, whereas that in $y$-periodic system (left) are coupled and gapped out when the bulk gap is too small. (d) The ratio $G/G^{(\infty)}$ of the tunneling conductance at finite width to that at infinite width vs the width $L$ when $\phi=13/21$. $t_x=1, t_y=0.7$ for all subgraph.}
\label{fig: edge}
\end{figure}

\textcolor{blue}{\em Edge transport.}--Now we turn to quantum transport contributed by edge states in a finite size system. %which is in sharp contrast in different directions.
The localization length $\xi_d$, characterizing the spatial distribution of the edge states along $d$-direction, may be dominated by two different aspects -- the topological gap $E_g$ or the quasiperiodicity.
 Similarly, we consider $t_x>t_y$. The edge states at $x$-boundary exhibit conventional behavior with $\xi_x \propto E_g^{-1}$, as the system is extended in the $x$-direction [Figs. \ref{fig: edge}(a,b)]. The localization length $\xi_y$ of edge states in the $y$-boundary may be dominated by the bulk gap $E_g$ or the AQP quantified by $q$. %Considering the case where $q\gg 1$, when $E_g$ is large ($E_g \gg q^{-1} t_x$), $\xi_y$ can still be characterized by the inverse of the gap, i.e., $\xi_y\propto E_g^{-1}$.
In most cases that $E_g$ is small ($E_g \ll q^{-1} t_x$), the universal scaling with exponential decay of the bulk versus $q$ in the $y$-direction determines that $\xi_y\propto q$, %will not reach $E_g^{-1}$,
and we analytically find~\cite{SM}
\begin{equation}
\xi_y=\left(-\log \frac{t_y}{t_x} + 2 q^{-1} \log 2 \right)^{-1},
\end{equation}
%given by $\xi_d= q/\log \lambda_d$, with $\lambda_d$ being a parameter in the ansatz $c_{k_{\bar{d}},j_d+1} = \lambda_d c_{k_{\bar{d}},j_d}$ that solves the Schr\"{o}dinger equation on the edge.
%{\color{red} The result means that the edge state wavefunction decays a factor of $\sim(t_y/t_x)^q$ moving into the bulk a unit cell, justifying the claim $\xi_y \sim q$.}
which is independent of $E_g$, as shown in Fig. \ref{fig: edge}(b).
 If we consider one direction having periodic boundary condition and another being open, the critical system size, beyond which the edge states in $x$(or $y$)-boundary are decoupled, is $L^c_{x}=2\xi_x\sim E_g^{-1}$, while
 $L^c_{y}=2\xi_y$ is significantly smaller than $E_g^{-1}$ for small gaps [Figs. \ref{fig: edge}(b,c)]). The longitudinal tunneling conductance between leads attached to two $d$ boundaries and through the 1D edge channels conducting along $\bar d$ boundaries reads (see ref.~\cite{SM}) $G_{xx}^{(L_y)}(G_{yy}^{(L_x)})\rightarrow G_{xx(yy)}^{(\infty)}=Me^2/h$ for sufficiently large $L_{y(x)}>L^c_{y(x)}$, where $M$ denotes the number of 1D edge channels contributed to the tunneling. Given that $L^{c}_{x}\gg L^c_{y}$ for the small gaps, when the system size $L_x=L_y=L$ with $L^c_{x}>L> L^c_{y}$, the edge states localized in the $x$-boundary ($y$-boundary) are coupled (decoupled), resulting in
 $G_{xx}^{(L)}$ (contributed by $y$-boundary edge states) being significantly larger than $G_{yy}^{(L)}$~\cite{SM}. %This is significantly different from the edge transport properties in previously studied Hall effect systems.
 %{\color{blue}We first consider $x$-direction having periodic boundary condition and $y$-direction being open. Then the critical length, beyond which the edge states in the $y$-direction are decoupled, is $L_{y,c}\approx 2q$[Fig. \ref{fig: edge}(c)]. So the edge transport characterized by the longitudinal tunneling conductance $G_{xx}^{(L_y)}$ with $E_F$ being in the gap, in the setup that is open in the $y$-direction and interfaces with metallic lead in the $x$($y$)-direction, satisfy $G_{xx}^{(L_y)}\rightarrow G_{xx}^{(\infty)}=Me^2/h$ for sufficiently large $L_y>L_{y,c}$, where $M$ denotes the number of edge state channels for the chosen bulk gap. Given that $L_{x,c}\gg L_{y,c}$ for the minigaps, then the edge transport described by $G_{xx}^{(L_y)}$ as above mentioned and the edge transport in another direction, described by $G_{yy}^{(L_x)}$ are substantially different~\cite{SM}.} So for $L_x \approx L_y \approx L$,  $G_{xx}^{(L)}$ is significantly larger than $G_{yy}^{(L)}$ for a broad region of $L$. For example, let $L=50$, which is a commonly accessible system size in various platforms, and $\phi=13/21$.
In summary, %$G_{yy}^{(L)}=G_{yy}^{(\infty)}$ when $E_F$ is located in the large gap, while
$G_{yy}^{(L)}<G_{yy}^{(\infty)}$ for the typical small gaps, but $G_{xx}^{(L)}=G_{xx}^{(\infty)}$ is independent of the gap size [Fig. \ref{fig: edge}(d)], rendering an exotic enhancement of $G_{xx}^{(L)}$ through edge states by AQP. This sharp contrast deeply reflects the different universal scalings (i.e. the power-law and exponential law) of the 2D bulk in $x$ and $y$ directions.
 %Then the only gap such that $G_{yy}^{(L)}=G_{yy}^{(\infty)}$ is when $E_F$ is located in the largest gap, whose corresponding Chern number $C=\pm 1$, with all other cases $G_{yy}^{(L)}<G^{(\infty)}$. In contrast, $G_{xx}^{(L)}=G_{xx}^{(\infty)}$ for all gaps, independent of the gap size [Fig. \ref{fig: edge}(d)]. Therefore, tunneling conductance of the edge transport in the $x$-direction is greatly enhanced by AQP.

\begin{figure}[!ht]
\centerline{\includegraphics[width=0.49\textwidth]{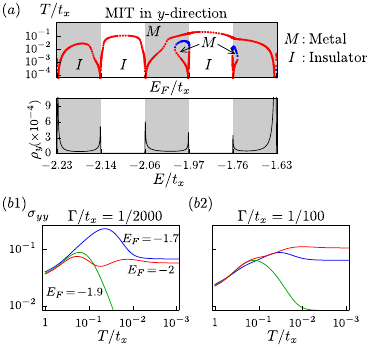}}
\caption{(a) Metal-insulator boundaries (upper panel) for small $\Gamma$ and the corresponding $\rho v_y^2$ (lower panel). Red (blue) dots denote the MIT temperature $T_0$. Gray (white) denotes the region of band (gap). Conductivity vs temperature for (b1) $\Gamma/t_x=1/2000$ and (b2) $\Gamma/t_x=1/100$. For all subgraphs, $t_x=1, t_y=0.7, \phi=8/13$.}
\label{fig: MIT}
\end{figure}

\textcolor{blue}{\em Finite-temperature anisotropic MITs.}--When finite temperature $T$ is considered, the system exhibits various MITs tuned by temperature with different $E_F$. %The most common MIT is described by the standard band theory, i.e. the system is in metallic(insulating) phase when $E_F$ is at the band(gap).
 The finite-temperature MITs are defined through $d\sigma/dT=0$ and can be interpreted using the Boltzmann equation, where the density of states $\rho$ and the square of the group velocity $v_d^2$ are key parameters.
%The coincidence of the foregoing definitions at $T=0$ is caused by the fact that the contribution to the conductivity, which is proportional to the product $\rho_d$ of the density of state and the group velocity in the $d$-direction according to Boltzmann formula, is greatest at the center of the band.
Transport in the $x$-direction is entirely determined by whether $E_F$ is located in the bulk band or in the gap, similar to previous results at zero temperature~\cite{SM}. In the $y$-direction, the system behaves as an insulator at sufficiently low $T$ for large $q$, except when $E_F$ is in the immediate vicinity of the band-gap boundary, where the product $\rho v_y^2$ is peaked [Fig. ~\ref{fig: MIT}(a)]~\cite{SM}. This feature is due to the near flat band configurations in the AQP regime, similar to the results in Fig.~\ref{fig: BulkT}(c2). We numerically find that the MIT in the $y$-direction approximately occurs at $T \sim \delta$, where $\delta$ is the energy difference between $E_F$ and the nearest band-gap boundary [Fig. \ref{fig: MIT}(b1, b2)]. In particular, the loop-like feature in Fig. \ref{fig: MIT}(a) indicates that multiple MITs with $T$ occur at specific Fermi energies, as shown for $E_F=-2$ in Fig. \ref{fig: MIT}(b1), as a consequence of the contribution from peaks of $\rho v_y^2$ at two band-gap boundaries (see Ref.~\cite{explainMITs} for a detailed explanation). %However, as seen from the red lines in Fig. \ref{fig: MIT}(b1) and (b2), increasing $\Gamma$ causes the multiple MITs to disappear. This is because, at certain $E_F$ for sufficiently small  (large) $\Gamma$, the transport can be dominated by more (fewer) peaks of $\rho v_y^2$ at different band boundaries.
In comparison, for large $T$ the system generally enters metal phase [Fig.~\ref{fig: MIT}(a)], similar to the Metal-I phase at large relaxation.

\textcolor{blue}{\em Discussion and conclusion.}-- %We have predicted novel universal transport phenomena, including the wave-packet dynamics, bulk and edge transport, in the 2D asymptotic quasiperiodic system which is characterized by a sequence of rational magnetic fluxes, but with an irrational limit. The results unveil multiple metal-insulator transitions driven by the interplay effect between asymptotic quasiperiodicity (AQP), relaxation, and finite temperature, and have a unified profound mechanism interpreted as the universal scaling laws emerging in the AQP regime, demonstrating a novel asymptotic connection to quantum phases in the quasiperiodic limit of the present anisotropic system.
We have predicted that the asymptotic quasiperiodicity (AQP) in a 2D anisotropic system host exotic universal wave-packet dynamics, bulk and edge transport phenomena, which have a unified profound mechanism interpreted as the universal scaling laws emerging in the AQP regime. The results unveil new metal-insulator transitions driven by the interplay effect between AQP, relaxation, and finite temperature, which demonstrate nontrivial asymptotic connection to quantum phases in the quasiperiodic limit. The prediction is feasible for experimental study.
%We have predicted that the asymptotic quasiperiodicity (AQP) in a 2D anisotropic system can exhibit exotic anisotropic transport phenomena in both bulk and edge, which demonstrate universal scaling behavior as $q$ varies, along with novel multiple MITs, and these results can be feasibly studied in experiments.}
The lattice with tunable magnetic flux can be experimentally achieved in various systems, such as cold atoms~\cite{RealizeHB,RealizeHB2,DMWeld}. The two main parameters $\Gamma$ and $T$ can be controlled. For cold atoms, the lowest temperature and relaxation coefficient can be tuned to about $10^{-2}$ and $10^{-4}$ of hopping energy, respectively~\cite{RelaxCA}. In solid platforms like quantum dots, these parameters can be as low as $10^{-3}$ and $10^{-4}$, respectively~\cite{RelaxQD, RelaxQD1, RelaxQD2}. With these parameters our predictions can be well observed by taking $\phi=8/13$ or $13/21$~\cite{notephi}. We point out that our predictions can be studied or directly applied to understand the transport phenomena in the recently highly attractive 2D Moir\'{e} materials, in which by adjusting the angle of rotation between the two layers, one can tune moir\'{e} materials to be periodic, quasiperiodic, or asymptotically quasiperiodic~\cite{Moire1, Moire2, Moire3, Moire4, Moire5, Moire6, Moire7, DEParker,WestonA,Magaud2012,Ahn2018}. Our work opens up an avenue to study the intriguing transport physics with AQP in high dimensions.

\begin{acknowledgments}
\textcolor{blue}{\em Acknowledgement.}--We thank Xin-Chi Zhou and Ming Gong for valuable discussions. This work was supported by National Key Research and Development Program of China (No. 2021YFA1400900 and No. 2022YFA1405800), the National Natural Science Foundation of China (Grants No. 12425401, No. 12261160368, No. 12104205, and No. 11921005), and the Innovation Program for Quantum Science and Technology (Grant No. 2021ZD0302000).

\textcolor{blue}{\em Note added.}--We recently became aware of a preprint~\cite{FLMoire} that similarly maps 2D strained Moir\'{e} systems with uniform magnetic field to an 1D Aubry-Andr\'{e}-Harper model to illustrate the anisotropic transport behavior. Apart from this mapping, our study focuses on the 2D asymptotic quasiperiodic regime, and discovers the novel universal transport phenomena, and the anisotropic MITs. Ref~\cite{FLMoire} further supports the applicability of our results to the 2D Moir\'{e} materials, including various types of lattice configurations such as honeycomb and triangular lattices.
\end{acknowledgments}

%\bibliographystyle{apsrev4-2}
%\bibliography{AnomalousTransport2Dv10}

\begin{thebibliography}{90}
\bibitem{AAnote}The Aubry-Andr\'{e}-Harper (AAH) model can be written as $H=\sum_j 2V_0 \cos (2\pi\phi j +\phi_0) c_j^\dagger c_j + t_0(c_j^\dagger c_{j+1} + h.c.)$. $V_0, t_0, \phi_0$ are respectively the onsite potential, hopping, and initial phase referred in the maintext. $\phi$ is irrational in the original AAH model and when it is rational, the model is referred as the ``periodic version of AAH model" in the main text. When $\phi$ is irrational, the critical point is at $V_0 = t_0$, so when $V_0$ is larger (smaller) than the critical value, the system is localized (extended).
\bibitem{explainMITs} Consider the Fermi energy $E_F=-2$, close to the peak located at the gap boundary $E=-1.97$ [see Fig. 4(a)]. Initially, as the temperature $T$ increases, the particles are excited to the states at the peak
$E=-1.97$, causing $\sigma_{yy}$ to increase and thus $d\sigma_{yy}/dT>0$, manifesting an insulator. After the states at the peak $E=-1.97$ are nearly fully occupied, $\sigma_{yy}$ will decrease with increasing $T$, turning the system into a metal with $d\sigma_{yy}/dT<0$. Further increasing $T$ leads to unoccupation of the states at the lower peak $E=-2.06$, which accordingly increases the hole contribution in this peak and enhance $\sigma_{yy}$ again, rendering an insulator with $d\sigma_{yy}/dT>0$. Finally, the system will reenter metal phase when $T$ increases to high enough, similar to the regime with large $\Gamma$ at zero temperature. The underlying mechanism of the multiple metal-insulator transitions around $E_F=-1.76$ can be similarly analyzed with the hole charge carrier picture. The absence of loop-like feature around $E=-2.06, -2.14$ is because the peaks of $\rho v_y^2$ at the two gap boundaries are large such that the metal phase at high $T$ extends to the very low $T$ regime.
\bibitem{notephi} We note that although we use the Fibonacci ratio $p_n/q_n$, our results are flexible. Indeed, our results are primiarily determined by $q_n$, as long as the ratio is finite and the limit is irrational. Then, for finite $n$ the value $p_n$ can vary within a certain range, such as $8/13$ being written as $7/13$ or $9/13$, without affecting our conclusions. This makes our predictions be of high feasibility for experimental realization.
\end{thebibliography}

\end{document}

%%%%%%%%%%%%%%%%%%%%%%%%%%%%%%%%%%%%%% %%   Supplementary Information %%%%%%%%%%%%%%%%%%%%%%%%%%%%%%%%%%%%%%
\renewcommand{\thesection}{S-\arabic{section}}
\setcounter{section}{0}  %  this will re-count section from 1
\renewcommand{\theequation}{S\arabic{equation}}
\setcounter{equation}{0}  %  this will re-count eq from 1
\renewcommand{\thefigure}{S\arabic{figure}}
\setcounter{figure}{0}  %  this will re-count eq from 1
\renewcommand{\thetable}{S\Roman{table}}
\setcounter{table}{0}  %  this will re-count eq from 1
\onecolumngrid \flushbottom %\onecolumn

\newpage

\begin{center}
\large \textbf{\large Supplementary Material: Anomalous universal quantum transport in 2D asymptotic quasiperiodic system}
\end{center}
In this supplementary material, we present (A) the numerical analysis of the bulk properties over $q$, and (B) their theoretical basis. (C) The zero-temperature phase diagram is shown for different values of $q$. We also include (D) the details of the longitudinal conductivity, with a consistency check of the related scaling laws, and (E) the transverse conductivity along with the related edge transport, including the Chern number of the gaps and an analytic proof of the localization length $\xi_y$. Then, we demonstrate that (F) in the $x$-direction, our system indeed behaves conventionally in the finite temperature regime, and we provide some details of the metal-insulator transition (MIT) in the $y$-direction. Finally, we discuss (G) the use of other quantities to characterize the localization and extension behavior of the wave function, and (H) the interesting physical phenomena at the isotropic point $t_x=t_y$.

\subsection{Numerical analysis of the scaling laws of bulk properties}
\begin{figure}[!ht]
\centerline{\includegraphics[width=15cm]{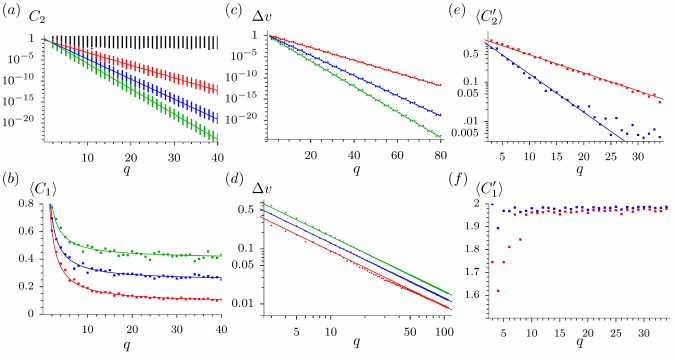}}
\caption{(a) $C_2$ vs $q$. The points denote data with varying $p$ and initial wavefunction $\psi(0)$. Solid lines represent the fitting of the mean $\langle C_2 \rangle \propto (t_y/t_x)^q$. Red, blue and green represent the data with $t_y/t_x=1/2,1/3$ and $1/4$ respectively. Black dots represent $C_1$, showing that $C_1$ only depends weakly on $q$. (b) $\langle C_1 \rangle$ vs $q$. Points and solid line represent the data and their fitted line respectively. Red, blue and green represent the data with $t_y/t_x=0.9,0.7$ and $0.5$, respectively. The general formula of the fitted lines is $0.807 (t_x - t_y) + 1.11 \sqrt{t_yt_x}/q$  (c) $\Delta v$ vs $q$ for $t_y<t_x$. Red, blue and green datas are calculated with $t_y/t_x = 1/2,1/3$ and $1/4$  respectively, showing $\Delta v \propto (t_y/t_x)^{q/2}$.  (d) $\Delta v$ vs $q$ for $t_y>t_x$. Red, blue and green dots and fitted lines represent $t_y/t_x = 5, 2$ and $10/7$ respectively, showing $\Delta v \propto q^{-1}\sqrt{t_xt_y}$. (e) $\langle C_2'\rangle$ vs $q$ for $t_y<t_x$. Red and blue dots and fitted lines represent $t_y/t_x = 0.9$ and $0.8$ respectively, showing $\langle C_2'\rangle \propto (t_x/t_y)^q$ (f) $\langle C_1'\rangle$ vs $q$ for $t_y>t_x$. Red and blue dots and fitted lines represent $t_y/t_x = 0.9^{-1}$ and $0.8^{-1}$ respectively, showing $\langle C_1'\rangle$ does not depend on $q$ for sufficiently large $q$.}
\label{GS1M}
\end{figure}

In this section, we present data concerning the numerical analysis of the scaling laws of different parameters of bulk transport with respect to the quantifier of asymptotic quasiperiodicity, $q$. As described in the main text, $C_2$ depends on $p$ and the initial wavefunction $\psi(0)$. For each $q$, we generate random values of $p$, which is coprime with $q$, and $\psi(0)$ to form the data points in Fig.~\ref{GS1M}(a). We note that the most significant correlation with $C_2$ is $q$, while $p$ and $\psi(0)$ have minimal effect, so it is natural to consider the average of $C_2$, $\langle C_2 \rangle$, over $p$ and $\psi(0)$. From the fitted lines, we establish that $\langle C_2 \rangle \sim t_x (t_y/t_x)^q$. Similarly, we also establish the power-law dependence on $q$, with $\langle C_1 \rangle \sim 0.807 (t_x - t_y) + 1.11q^{-1}\sqrt{t_yt_x}$ [Fig.~\ref{GS1M}(b)].

The average group velocity, whose exact formula is not shown in the main text, is defined as $\tilde{v}_{d} (q) = g(q)^{-1} \sum_{m,p} |v_{k_{xM},k_{yM}}^{(m,p/q,d)}|$, where $v_{k_{xM},k_{yM}}^{(m,p/q,d)}$ is the group velocity of the $m$th band in the $d$-direction of the system at $\phi=p/q$, with $k_x = k_{xM}, k_y = k_{yM}$. Here, $k_{xM}$ and $k_{yM}$ denote the quasi-momentum that corresponds to the  maximum group velocity. The normalization constant is given by $g(q) = q \phi(q)$, where $\phi(q)$ is the number of integers less than $q$ that are coprime with $q$. According to~\cite{totalbandwidth}, the total bandwidth of the system is $4(t_y - t_x)$ for infinite $q$ and $t_y>t_x$. Therefore, the mean group velocity is $4(t_y - t_x)/q \times (\pi/q)^{-1} = 4/\pi \times (t_y - t_x) $, since the system is periodic when $k\rightarrow 2\pi/q + k$. Similar to the above numerical analysis, we averaged the velocties over $p$ and bands, and recovered the numerical relation between $\Delta v = \tilde{v}_y - \tilde{v}_y^{(\infty)}$ and $q$ [Fig.~\ref{GS1M}(c,d)].

\begin{figure}[!ht]
\centerline{\includegraphics[scale=1.2]{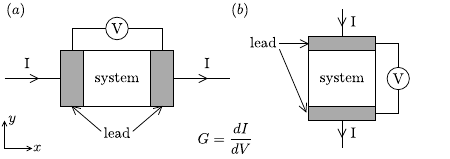}}
\caption{(a,b) Schematic diagrams showing how the tunneling conductance $G_{xx(yy)}$ is obtained from the current $I$ and the voltage $V$. }
\label{GS5M}
\end{figure}

Lastly, we propose employing the real-space longitudinal tunneling conductance $G_{xx(yy)}$ in the $x(y)$ direction to detect bulk transport.  This conductance is calculated numerically using non-equilibrium Green's function in a  real-space system of size $L_y(L_x)$ with two leads at the ends [Fig.~\ref{GS5M}]. We find that the conductance $G_{xx(yy)}$ in the $x$($y$) direction satisfies
\begin{equation}
G_{xx} = C_1' L_y; G_{yy} = \begin{cases}
C_2' L_x & \text{for rational }\phi \\
C_3' & \text{for irrrational }\phi
\end{cases},
\end{equation}
where numerical analysis shows that $\langle C_1'\rangle$ is of order $1$ and that $\langle C_2'\rangle \sim \widetilde{C}_2' (t_y/t_x)^q$[Fig.~\ref{GS1M}(e,f)]. Here $C_{1(2)}'$ depends on $p$ and Fermi energy $E_F$. For each $q$, we compute the average of $G_{xx}/L_y$ or $G_{yy}/L_x$ over $p$ and over random $E_F$ that cuts through the bulk. Notably, for $G_{xx}$ and $G_{yy}$ in the case of rational $\phi$, the proportionality of the system size indicates the presence of bulk transport in the system. The similar exponential dependence on $q$ suggests a comparable role of $q$ in the calculation of real-space conductance.

\subsection{Theoretic basis of the scaling law of $\langle C_2 \rangle$ and $\Delta v$}
In this section, we discuss the theoretical basis for the exponential decay of $\langle C_2 \rangle$ and $\Delta v$ with respect to $q$, as well as the additional $1/2$ factor in the exponent of the scaling of $\Delta v$. We begin by showing the similarity between Hamiltonians when Fourier-transformed in a single direction versus when Fourier-transformed in two directions. When we choose the $A_d$ gauge, the Hamiltonian can be Fourier transformed in the $d$-direction, i.e.,
\begin{eqnarray}
H &=& \sum_{k_x} H_\text{1D}(k_x) = \sum_{k_x,j_y} 2t_x \cos (k_x-2\pi\phi j_y) c_{k_xj_y}^\dagger c_{k_xj_y} + (t_y c_{k_x,j_y+1}^\dagger c_{k_xj_y}+h.c.)  \quad {\rm for}\: A_x\:{\rm gauge}, \nonumber \\
H &=& \sum_{k_y} H_\text{1D}(k_y) = \sum_{k_y,j_x} 2t_y \cos (k_y+2\pi\phi j_x) c_{j_xk_y}^\dagger c_{j_xk_y} + (t_x c_{j_x+1,k_y}^\dagger c_{j_xk_y}+h.c.)  \quad {\rm for}\: A_y\:{\rm gauge}.
\end{eqnarray}
For rational $\phi$, the Hamiltonian can be further Fourier transformed in the $\bar{d}$-direction, i.e.,
\begin{eqnarray}
H &=& \sum_{k_xk_y} \mathcal{H}_{A_x}(k_x,k_y) = \sum_{k_x,k_y}2t_y\cos(k_y) c_{k_xk_y}^\dagger c_{k_xk_y} + (t_x e^{ik_x} c_{k_xk_y}^\dagger c_{k_x,k_y-2\pi\phi}+h.c.))  \quad {\rm for}\: A_x\:{\rm gauge},\nonumber \\
H &=& \sum_{k_xk_y} \mathcal{H}_{A_y}(k_x,k_y) = \sum_{k_x,k_y}2t_x\cos(k_x) c_{k_xk_y}^\dagger c_{k_xk_y} + (t_y e^{ik_y} c_{k_xk_y}^\dagger c_{k_x+2\pi\phi,k_y}+h.c.))  \quad {\rm for}\: A_y\:{\rm gauge},
\end{eqnarray}
which can be rewritten into a form resembling the periodic AAH model as mentioned in the main text:
\begin{eqnarray}
\mathcal{H}_{A_x} &=& \Psi_{k_x,\phi}^\dagger [2t_y D_\phi(k_y) + (t_x e^{-ik_x} M_1 + h.c.)] \Psi_{k_x,\phi}, \nonumber\\
\mathcal{H}_{A_y} &=& \Psi_{-\phi,k_y}^\dagger [2t_x D_{-\phi}(k_x) + (t_y e^{-ik_y} M_1 + h.c.)] \Psi_{-\phi,k_y},
\end{eqnarray}
where $\phi=p/q$, $2\pi\phi$ is the momentum spacing, $\Psi_{k_x,\phi} = (c_{k_xk_y}, c_{k_x,k_y-2\pi\phi},\cdots,c_{k_x,k_y-2\pi(q-1)\phi})^T$, $\Psi_{-\phi,k_y} = (c_{k_xk_y}, c_{k_x+2\pi\phi,k_y},\cdots,c_{k_x+2\pi(q-1)\phi,k_y})^T$, $D_\phi(k) = \text{diag} (\cos k,\cos( k-2\pi\phi),\cdots,\cos [k-2\pi(q-1)\phi])$ and $M_1$ is a circulant matrix with $c_1 = 1$, meaning all elements on the lower diagonal and at the top-right corner are $1$, with all others being $0$. We note that the Hamiltonians $\mathcal{H}_{A_x}$ and $\mathcal{H}_{A_y}$ are explicitly related by a gauge transformation $\mathcal{H}_{A_x} = U^\dagger \mathcal{H}_{A_y} U$, where $U = (v_0, v_1, \cdots, v_{q-1})$ and $v_n = (1, e^{i2\pi n\phi}, e^{i4\pi n\phi}, \cdots, e^{i2(q-1)\pi n\phi})^T$. Therefore, in the main text, we simplify the notation by omitting the subscript in $\mathcal{H}$ in the corresponding expression.
%The Hamiltonian $H_x''$ and $H_y''$ are related explicitly by a gauge transformation $H_x'' = U^\dagger H_y'' U$, where $U = (v_0, v_1, \cdots, v_{q-1})$ and $v_n = (1, e^{i2\pi n\phi}, e^{i4\pi n\phi}, \cdots, e^{i2(q-1)\pi n\phi})^T$. Such observation simplifies the below discussion, but still keep its generality, since virtually all kinds of periodic function translate similarly from the singly Fourier transformed Hamiltonian to the doubly Fourier transformed Hamiltonian. 
The $\cos$ function in $H_{\text{1D}}$ with the $A_x$ gauge indicates that the eigenfunction decays exponentially in the $y$-direction for $t_y < t_x$, as the geometric mean of its diagonal terms satisfies
\begin{equation}
\GeoMean\limits_{k=0}^{2\pi} 2t_x(\cos k-r) \defeq \left(\prod_0^{2\pi} \left|2 t_x (\cos k - r)\right|^{dk}\right)^{1/(2\pi)} \defeq \exp\left[\frac{1}{2\pi}\int_0^{2\pi}dk\log\left|2t_x(\cos k-r)\right|\right]= t_x
\end{equation}
for $1>r>-1$. Therefore, when $t_y<t_x$ the off-diagonal terms can be treated as perturbation for sufficiently large $q$, where the continuous integral accurately represents the discrete geometric mean. Thus, the effective hopping of a charge carrier across a full unit cell is proportional to $t_{\text{eff}} = t_x (t_y/t_x)^q$, which matches the scaling of $\langle C_2 \rangle$.

Secondly, the maximum velocity in the $y$-direction, $v_y$, of each band actually corresponds to the maximum velocity of $\mathcal{H}_{A_y}$ over $k_x$ and $k_y$. This is proportional to the maximum bandwidth of the $\mathcal{H}_{A_y}$ gauge when treating $k_x$ as a parameter and $k_y$ as quasi-momentum. To achieve maximal bandwidth, $k_x$ should be tuned so that some diagonal terms of $\mathcal{H}_{A_y}$ are equal, thereby minimizing the perturbation effect of the off-diagonal terms and maximizing the bandwidth in the $y$-direction. For example, if $\phi=5/8$ and $k_x = 0$, then the second and sixth diagonal terms of $\mathcal{H}_{A_y}$ are the same, i.e., $2t_x\cos \left(2\times5\pi/4\right) =2t_x\cos \left(6\times5\pi/4\right) = 0$. Thus, the charge carrier with maximum transport power hops from the second to the sixth site, and then to the second site in the next unit cell, completing a period. This motion results in the bandwidth and the maximum velocities being of the order of $(t_y/t_x)^4 = (t_y/t_x)^{q/2}$. To generalize this result, we assume the diagonal terms to be $V_i(k_x)$, where $i$ indicates the $i$th diagonal term, and the geometric mean of $V$ is $1$ (which can be adjusted by absorbing extra factors into $t_x$). There are a total of $q$ diagonal terms, with the index $i$ having a period of $q$. If for some $k_x$, $V_{i_1}(k_x) = V_{i_2}(k_x) = \cdots = V_{i_n}(k_x) = V_{i_1+q}(k_x) = V_0$ with $i_1+q\geq i_n\geq \cdots\geq i_1$, then the bandwidth $E_{b,y}$ in the $y$-direction of the bands with $E \approx V_0$ satisfies
\begin{equation}
E_{b,y} = \gamma \times \begin{cases}
t_x (t_y/t_x)^{M_{n/2} - m_{n/2-1}} & \text{for }n\text{ is even} \\
t_x (t_y/t_x)^{q-2m_{(n-1)/2}} & \text{for }n\text{ is odd},
\end{cases}
\end{equation}
where $\gamma$ is some parameter of order 1. $M_j$ ($m_j$) represents the maximum (minimum) of the sums of $j$ items chosen from the set $ \{z_1 = i_2-i_1,  z_2 = i_3-i_2, \dots, z_n = i_1+q-i_n\}$. The chosen items $z_{l_1}, z_{l_2}, \dots, z_{l_j}$ must satisfy the following conditions: (i) $|l_\alpha -l_\beta| > 1$ for all $1 \leq \alpha, \beta \leq j$, (ii) if
$l_\alpha = 1$, then $l_\beta \neq n$ for all $\beta$, and (iii) if $l_\alpha = n$, then $l_\beta \neq 1$ for all $\beta$.
For example, when $n=7$, $M_3 =\max\{z_1+z_3+z_5, z_1+z_3+z_6, z_1+z_4+z_6, z_2+z_4+z_6, z_2+z_4+z_7, z_2+z_5+z_7, z_3+z_5+z_7\}$, and $m_2 = \min\{z_1+z_3, z_1+z_4, z_1+z_5, z_1+z_6, z_2+z_4, z_2+z_5,  z_2+z_6, z_2+z_7, z_3+z_5, z_3+z_6, z_3+z_7, z_4+z_6, z_4+z_7, z_5+z_7\}$. Thus, the average of the maximal bandwidth is primarily determined by the case where $z_1 = z_2 = \cdots = z_n = q/n$, and then
\begin{equation}
\tilde{v}_y \sim t_x (t_y/t_x)^{q/n},
\end{equation}
where $n=2$ for the $\cos$ potential considered in our work.

\subsection{Zero temperature phase diagram for different $q$}
\begin{figure}[!ht]
\centerline{\includegraphics{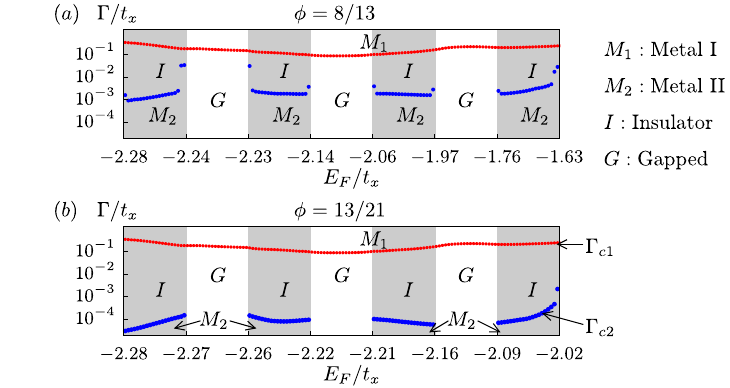}}
\caption{Phase diagrams. The parameters chosen are $t_y=0.7$ and $t_x=1$ for the cases (a) $\phi = 8/13$ and (b) $\phi=13/21$. (b) clearly shows a smaller $M_2$ region and a larger $I$ region compared to (a).}
\label{GS2M}
\end{figure}
In this section, we present the generic phase diagrams at zero temperature for $\phi = 8/13$ and $\phi=13/21$ [Fig.~\ref{GS2M}(a,b)]. We note that for $\phi=13/21$, the region $M_2$ is reduced, consistent with the discussion in the main text. It is also noteworthy that $\Gamma_{c1}$ is essentially independent of $\phi$, whereas $\Gamma_{c2}$ is reduced by a factor of $10$ to $30$ when $q$ changes from $13$ to $21$, consistent with the scaling law $\Gamma_{c2}/t_x \sim (t_y/t_x)^q$.

\subsection{Details in longitudinal conductivity}
\begin{figure}
\centerline{\includegraphics{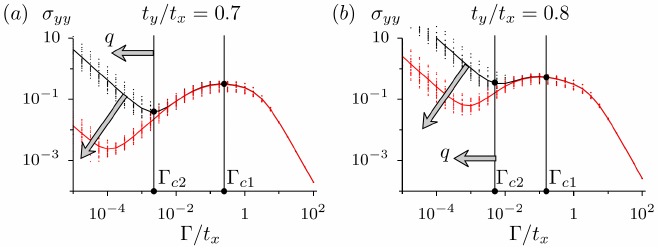}}\caption{Generic behavior of the longitudinal conductivity $\sigma_{yy}$ at different Fermi energies located in the bulk vs $\Gamma$. The dots represent Fermi energies chosen at the centers, as well as the upper and lower quartiles of all bands, showing similar scaling properties. The lines represent the mean of the calculated $\sigma_{yy}$. Black (red) represents the data for $\phi = 8/13$ ($13/21$). }
\label{GS6M}
\end{figure}
In this section, we discuss the details of the calculation of the Kubo formula when applied to the longitudinal conductivity. We first introduce the general Kubo formula at zero temperature:
\begin{eqnarray}
\sigma_{dd'} &=& \lim_ {\omega\rightarrow 0} \frac{ie^2}{\omega \hbar} \int \frac{dk_0}{2\pi}\frac{d^d k}{\left(2\pi\right)^d} \text{ tr}\left[\left(\left. G_{ik_0+iq_0}\right|_ {iq_0 \rightarrow \omega+i 0^+}-G_{ik_0}\right) \frac{d\mathcal{H}}{dk_{d'}} G_{ik_0} \frac{d\mathcal{H}}{dk_d}\right]\\
&=& \frac{e^2}{h} \int \frac{d^d k}{(2\pi)^d} \text{ tr}\left[\left(G^+_{0}-G^-_{0}\right) \frac{d\mathcal{H}}{dk_{d'}} G^-_{0} \frac{d\mathcal{H}}{dk_d} - i\int dk_0 \left(G^{\text{sgn}(k_0)}_{ik_0}\right)^2 \frac{d\mathcal{H}}{dk_{d'}} G^{\text{sgn}(k_0)}_{ik_0} \frac{d\mathcal{H}}{dk_d}\right], \label{kubooriginal}
\end{eqnarray}
where $G_{ik_0} = \left(ik_0+\mu+i\Gamma \text{sgn }{k_0} - \mathcal{H}\right)^{-1}$, $G_{ik_0}^{\pm} = \left(ik_0+\mu\pm i\Gamma- \mathcal{H}\right)^{-1}$ and $\Gamma$ is the finite relaxation rate, assumed to be independent of momentum and band. When $d=d'$, the above expression simplifies to 
\begin{equation}
\sigma_{dd} = \frac{e^2}{h} \times \mathfrak{R}\left\{\int \frac{d^2 k}{(2\pi)^2} \text{ tr}\left[\left(G^+_{0} - G^-_{0}\right)\frac{d\mathcal{H}}{dk_d} G^-_{0} \frac{d\mathcal{H}}{dk_d}\right]\right\},
\end{equation}
where $\mathfrak{R}$ represents the real part of the given expression.
The scaling of the conductivity $\sigma_{yy}$ in the three universal regions, metal I, metal II and insulator, follows $\sigma_{yy}\sim\Gamma^{-1,1,-2}$, respectively. We provide a more detailed numerical analysis below. As shown in Fig.~\ref{GS6M}, we calculate $\sigma_{yy}$ using multiple Fermi energies to support our claim about the scaling in the insulating region and the transition points. We further verify this claim by calculating the asymptotic behaviors for both $\Gamma\rightarrow 0$ and $\Gamma\rightarrow\infty$. Specifically, for $\Gamma\rightarrow\infty$,
\begin{eqnarray}
\sigma_{yy} \left(\Gamma\rightarrow\infty\right) &=& \frac{2e^2}{h\Gamma^2}\int \frac{dk_x}{2\pi}\frac{dk_y}{2\pi} \text{ tr}\left(\frac{d\mathcal{H}}{dk_y} \frac{d\mathcal{H}}{dk_y}\right) = 4t_y^2 \Gamma^{-2}.
\end{eqnarray}
For $\Gamma \rightarrow 0$,
\begin{eqnarray}
\sigma_{yy} \left(\Gamma\rightarrow 0\right)  &=& \frac{e^2}{h\Gamma} \int \frac{dk_x}{2\pi}\frac{dk_y}{2\pi} \sum_b \delta\left(\epsilon_b(k_x,k_y)-\mu\right)\left(\frac{\partial\epsilon_b}{\partial k_\alpha}\right)^2 \nonumber\\
&=& \frac{e^2}{h\Gamma}\frac{q}{\pi} \times \left| \mathfrak{I}\left\{\left(\frac{df}{d\mu}\right)^{-1}\sqrt{\frac{1}{4t_x^{2q}-8t_x^qt_y^q+4t_y^{2q}-f^2}}\left[\left(4t_x^{2q}-8t_x^qt_y^q+4t_y^{2q}-f^2\right)E\left(\mathfrak{F}_{f,t_x,t_y,q}\right)\phantom{\frac{2t_x^q}{2t_x^q}}\right.\right.\right. \nonumber\\
&& \left.\left.\left. -2\left(2t_y^q-f\right)\left(2t_y^q-2t_x^q+f\right)K\left(\mathfrak{F}_{f,t_x,t_y,q}\right)+8t_y^q f  \Pi \left(\frac{2t_x^q+2t_y^q+f}{2t_y^q-2t_x^q+f},\mathfrak{F}_{f,t_x,t_y,q}\right)
\right]\right\}\right|\nonumber\\
&=& \frac{e^2t_x}{h\Gamma}\frac{q}{\pi}\times \left| \mathfrak{I}\left\{\left(\frac{df'}{d\mu'}\right)^{-1}\sqrt{\frac{1}{(1-r)^2-f'^2}}\left[\left((1-r)^2-f'^2\right)E\left(\mathfrak{F}_{f,t_x,t_y,q}\right)\phantom{\frac{2t_x^q}{2t_x^q}}\right.\right.\right. \nonumber\\
&& \left.\left.\left. -2\left(r-f'\right)\left(r-1+f'\right)K\left(\mathfrak{F}_{f,t_x,t_y,q}\right)+f'  \Pi \left(1+\frac{2r}{r-1+f'},\mathfrak{F}_{f,t_x,t_y,q}\right)
\right]\right\}\right|\nonumber\\
&=& \frac{e^2t_x}{h\Gamma} \frac{qr^2}{2\sqrt{1-f'^2}}\left(\frac{df'}{d\mu'}\right)^{-1} + O(r^3),
\end{eqnarray}
where $\epsilon_b$ is the energy of the $b$th band, $\mathfrak{F}_{f,t_x,t_y,q} = \left(f^2-4t_x^{2q}-8t_x^qt_y^q-4t_y^{2q}\right)/ \left(f^2-4t_x^{2q}+8t_x^qt_y^q-4t_y^{2q}\right)$, $f = \text{det}\left(\mathcal{H}(\pi/2q,\pi/2q) - \mu I \right)$, $f' = f/t_x^q$, $\mu' = \mu/t_x$ and $r=t_y^q/t_x^q$, which is small when $t_y<t_x$ and $q\gg 1$. $E, K$ and $\Pi$ are the elliptic functions of the first, second and third kinds, respectively. The quantity $df'/d\mu'$ is of the order of $q$, so in this limit, $\sigma \sim (t_y/t_x)^{2q}t_x\Gamma^{-1}$. A consistency check can be performed by comparing the five scalings behaviors: three from the regions of $\sigma$ and two from the critical relaxation rates $\Gamma_{c1,c2}$,
\begin{equation}
\sigma_{yy,M_1} \sim 4t_y^2 \Gamma^{-2}, \sigma_{yy,M_2} \sim t_x (t_x/t_y)^{2q} \Gamma^{-1}, \sigma_{yy,I} \sim \Gamma, \Gamma_{c1} \sim t_y, \Gamma_{c2} \sim t_x (t_x/t_y)^q.
\end{equation}
Specifically, the consistency can be observed by noting that both $\Gamma_{c1}$ and $\sigma_{yy,M_1}$ are independent of $q$, therefore $\sigma_{yy,I}$ is also independent of $q$. Then, by comparing the scaling of $\sigma_{yy,M_2}$ and $\sigma_{yy,I}$, one can independently approximate $\Gamma_{c2}$, which is consistent with the result of the numerical analysis, further verifying the numerically derived scalings.

\subsection{Details in transverse conductivity and edge transport}
\begin{figure}[!ht]
\centerline{\includegraphics{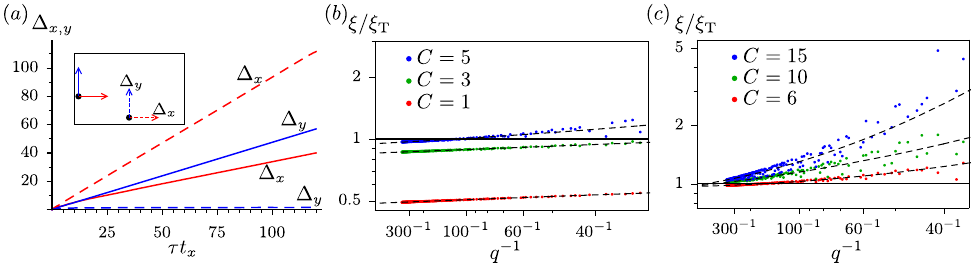}}
\caption{(a) DMSDs vs time. Solid (dashed) lines represent DMSDs when the initial position is at the $x$-boundary (or $y$-boundary). Red (blue) lines represent DMSDs in the $x$-direction (or $y$-direction). $\phi = (\sqrt{5}-1)/2$. (b,c) The ratio of the numerically obtained $w$ to the theoretically obtained $w_T$. The energies across $q$ are chosen at the centers of gaps where the Chern number remains the same. $p$ is chosen such that $p/q$ is closest to  $(\sqrt{5}-1)/2$. }
\label{GS3M}
\end{figure}
In this section, we discuss the wavepacket dynamics at the edge, the transverse conductivity, and the analytic derivation of the localization length of edge states. For the wavepacket dynamics, we consider DMSDs with the initial position $r_0$ at the edge. The transport behavior when $r_0$ is located at the $x$-boundary is qualitatively different from that at the $y$-boundary: Both $f_x$ and $f_y$ is diffusive when $r_0$ is at the $x$-boundary, while $f_x$ is diffusive and $f_y$ is localized when $r_0$ is located at the $y$-boundary, as shown in Fig. \ref{GS3M}(a) The implication is that the topological edge states are unaffected by the incommensurate flux $\phi$, even when the bulk is localized in one direction. Thus, DMSDs along the edge indicate that the edge is always diffusive, while the DMSDs perpendicular to the edge are still influenced by the bulk displacement function.

Regarding the transverse conductivity $\sigma_{xy}$, according to the Kubo's formula given in the main text, it simplifies in the limit as $\Gamma \rightarrow 0$:
\begin{equation}
\sigma_{xy} = \frac{e^2}{h} \int \frac{dk_xdk_y}{\left(2\pi\right)^2} \mathfrak{I}\left[\sum_{nm}\frac{\left(\psi_n^* \frac{d\mathcal{H}}{dk_x} \psi_m \psi_m^* \frac{d\mathcal{H}}{dk_y} \psi_n\right)}{(E_n-E_m)^2} \right],
\end{equation}
where $\psi_{n(m)}$ denotes the wavefunction of the filled (unfilled) bands. When the Fermi level is located at the gap, the result is proportional to the Chern number of the bands involved. For an arbitrary rational $\phi$, the Chern number of the $n$th band, $C_{p/q}^{(n)}$, is
\begin{equation} \label{Chern}
C_{p/q}^{(n)} = \left(p^{-1}_{\text{mod }q}n(1-\delta_{n,q/2})+\floor{\frac{q}{2}}\right)_{\text{mod }q} - \left(p^{-1}_{\text{mod }q}(n-1)(1-\delta_{n-1,q/2})+\floor{\frac{q}{2}}\right)_{\text{mod }q},
\end{equation}
where the subscript $\text{mod } q$ indicates that the number is calculated in the field of $\mathbb{Z}/q\mathbb{Z}$ (for example, $3^{-1}_{\text{mod }13} = 9$). Therefore, for all rational $\phi$, the edge states continue to exist.

However, even when the edge states exist, their wavefunctions show qualitative differences. For the edge along $y$, the edge state penetrates into the bulk with a localization length proportional to the gap width $E_g$, whereas for the edge along $x$, the localization length of the edge state does not depend on the gap width. We can see this analytically through the following calculation. To begin, we note that the system can be viewed as having a unit cell with $q$ sites, so we can calculate the edge wavefunction using the ansatz $c_{i\alpha} = \lambda^i c_{0\alpha}$, where $\alpha = 0,1,\dots,q-1$. Then the parameter $\lambda$ satsifies $\det\left(-B(k)+\lambda\left(E-H(k)-\lambda B^\dagger(k)\right)\right) = 0$, where $B(k)$ is a $q\times q$ matrix describing the hopping from $i$ to $i+1$, and $H(k)$ is a matrix describing the onsite potential. Now, for the $x$-boundary
\begin{eqnarray}
\lambda_{\pm} &=& \frac{t_x^{-q}}{2} \left((-1)^q f_{p,q} - 2t_y^q \cos(k_yq)\pm\sqrt{f_{p,q}^2-4t_x^{2q}+4t_y^q\cos(k_yq)\left(t_y^q\cos(k_yq)-(-1)^qf_{p,q}\right)}\right) \\
&=& \tilde{f} \pm \sqrt{\tilde{f}^2-1},
\end{eqnarray}
where $f_{p,q} = f_{p,q}(e,t_x,t_y)$ is the part of $\det(\mathcal{H}-e I)$ that is independent of $k_x$ and $k_y$, and $\tilde{f} = t_x^{-q}/2 \times \left((-1)^q f_{p,q}-2t_y^q \cos(k_yq)\right)$. Since $f$ is a continuous function of $e$, for a sufficiently small gap, the value of $f$ can be approximated by $f_{t/b}$, the value at the top or bottom of the band. Since the top or bottom of the band occurs only at $(k_x,k_y) = (n\pi/q,m\pi/q)$ (where $n$ and $m$ are integers), and $\det(\mathcal{H}-eI)=0$ at that point, $f_{t/b} = \pm 2\left(t_x^q+t_y^q\right)$. If $t_y>t_x$, as $q \rightarrow \infty$, $\lambda_\pm \approx (2\tilde{f})^{\pm 1} \approx \left[2(t_y/t_x)^q \times ((-1)^{i_g}-\cos(k_yq))\right]^{\pm 1}$ ($i_g$ denotes the number of bands that are below the gap) except when $((-1)^{i_g}-\cos(k_yq))=0$. In this case, the localization length of the edge state is given by $\xi=|q(\log |\lambda|)^{-1}| \approx \log (t_y/t_x)^{-1}$. A more careful analysis yields the next-order term, which is included in the main text. Thus in this case, except at some particular $k_y$, the edge state rapidly decays no matter how small the gap is. On the other hand, if $t_x>t_y$, as $q\rightarrow \infty$, $f \approx (-1)^{i_g}\left(1+2a_{p,q,g}^2\left(E_g^2/4-(e-E_{i_g,c})^2\right)\right)$, where $E_{i_g,c}$ is the energy of the center of the $i_g$-th gap, and $a_{p,q,i_g}$ is a positive constant. Therefore, at the center of the gap, $\lambda_{i_g,c,\pm} \approx (-1)^{i_g}\left(1\pm\sqrt{4a_{p,q,i_g}^2\left(E_g^2/4-(e-E_{i_g,c})^2\right)}\right) = (-1)^{i_g}\left(1\pm E_g a_{p,q,i_g}\right)$, and $\xi\approx (a_{p,q,i_g}E_g)^{-1}$. In this case, the localization length of the edge state is inversely proportional to the gap. We also compared the numerically obtained $w$ with the theoretically obtained $w_T$ and found that as long as the gap is small enough (i.e. large Chern number) and $q$ is large enough, the theoretic and numerical results coincides  [Fig.~\ref{GS3M}(b,c)].

\subsection{Finite temperature phase diagram in both directions}
\begin{figure}[!ht]
\centerline{\includegraphics{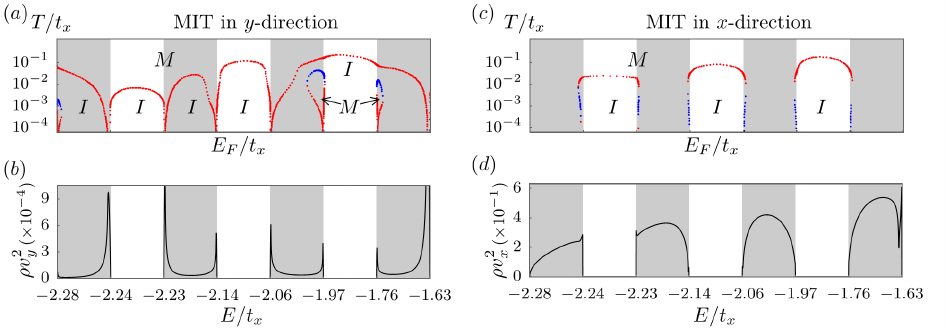}}
\caption{(a) The phase diagram classified by $\sigma_{yy}$. (b) The product of density of states (DOS) and $y$-velocity. (c) The phase diagram classified by $\sigma_{xx}$. (d) The product of DOS and $x$-velocity. Parameters chosen: $t_x = 1, t_y =0.7, \Gamma = 1/100, \phi=8/13$.}
\label{GS4M}
\end{figure}
In this section, we also present the metal-insulator transition (MIT) in the $x$-direction and provide some details of the MIT in the $y$-direction. The MIT in the $x$-direction resembles that at $T=0$, i.e. for sufficiently low temperatures, the system is insulating/metallic when the Fermi energy is in the gap/band [Fig.~\ref{GS4M}(c)]. This behavior indeed originates from the fact that the product of DOS and $x$-velocity is maximal at the center of the band [Fig.~\ref{GS4M}(d)]. For comparison, we also present a similar diagram in the $y$-direction.

In the main text, we also mention that the system is insulating in the $y$-direction for sufficiently large $q$ and for almost all $E_F$, except in the immediate vicinity of the bulk-gap boundary. Specifically, for $q=13$, the total range of such exceptional $E_F$ is of the order of $10^{-3}$, compared to the total energy range of $5.56$.

\subsection{Characterization of the bulk state wavefunctions}
\begin{figure}
	\centerline{\includegraphics{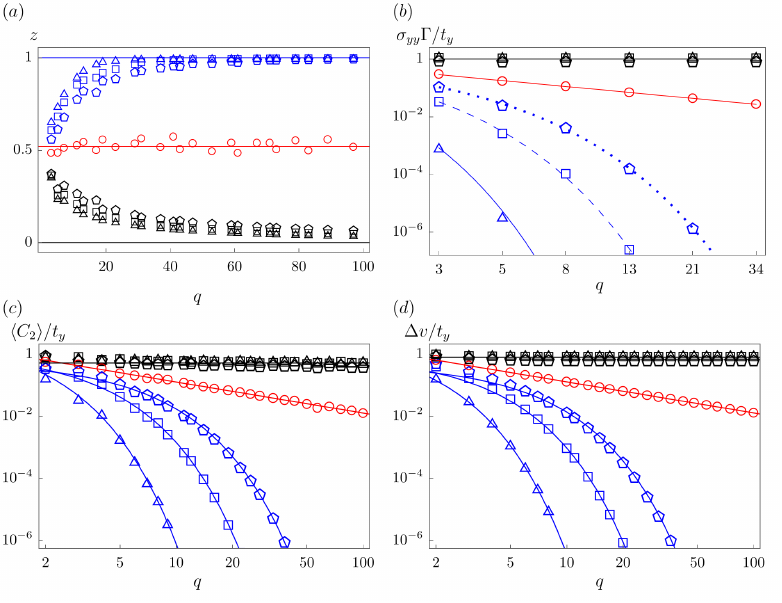}}\caption{In all subgraphs, $t_x = 1$. The black pentagon, square and triangle represent $t_y=4$, $t_y=3$ and $t_y=2$, respectively. The red circle represents $t_y=1$. The blue pentagon, square and triangle represent $t_y=0.7$, $t_y=0.5$ and $t_y=0.2$, respectively. The black, red, blue lines indicate asymptotic behavior or fitted scaling in the extended, critical and localized regimes, respectively. (a) $z$ vs $q$: $z\rightarrow 1$ in the localized regime (blue line) and $z\rightarrow 0$ in the extended regime (black line). In the critical regime, $z \rightarrow \kappa \approx 0.52$ as indicated by the red line. (b) The conductivity $\sigma_{yy}\Gamma/t_y$ vs $q$. (c) $\langle C_2 \rangle$ vs $q$. (d) $\Delta v$ vs $q$.  (b-d) In the localized regime, the blue fitted line indicates an exponentially decaying tendency as $q$ increases, while the red fitted line indicates an algebraically decaying tendency. The black fitted line indicates independence of $q$. }
	\label{GS7M}
\end{figure}

Other quantities can also be used to characterize the localization and extension behavior of the wave function. We take periodic boundary conditions, choose the gauge $\vec A=2\pi\phi j_y\hat e_x$, set $L_y=q$, and introduce a dimensionless quantity~\cite{RSlength}
\begin{equation}
\tilde{z} = \sum_{j_xj_y} e^{i2\pi j_y/q} \left| \psi_{j_xj_y} \right|^2.
\end{equation}
In the extreme localized state (with $\left|\psi_{j_xj_y}\right|^2 \sim \delta_{j_xj_{x0},j_yj_{y0}}$), $\left|\tilde{z}\right| \rightarrow 1$, and in the extreme extended state (with $\left|\psi_{j_xj_y}\right|^2 \sim (L_xL_y)^{-1}$) $, \left|\tilde{z}\right| \rightarrow 0$. We numerically find that the quantity $\tilde{z}$ is approximately independent of $p$ and $L_x$, except for certain choices of $p$, $q$, and $L_y$ where the system is highly degenerate, making $\tilde{z}$ not well-defined since the eigenstates can be freely mixed within the degenerate subspaces.
To avoid the numerical errors from high degeneracy, we define the mean of $\tilde{z}$ as $z = \mathcal{N}^{-1}\sum_{p,L_x,m} \left|\tilde{z}\right|(p,L_x,m)$ to quantify the localization behavior of the whole system as a function of $q$, where $\mathcal{N}$ is the total number of selected eigenstates ($m$ is the eigenstate index) and the total number of selected samples with different $p$ and $L_x$. The mean $z$ is not affected by those $p$ and $L_x$, for which the system is highly degenerate. For $q \rightarrow \infty$, we can see that $z \rightarrow 0$ for $t_x>t_y$, indicating the states along $x$ direction to be extended, and $z \rightarrow 1$ for $t_x<t_y$, indicating the states along $x$ direction to be localized [see Fig.~\ref{GS7M}(a)]. This is consistent with the results that we obtain using other quantities. On the other hand, one can see that $z \rightarrow \kappa \approx 0.52$ for $t_x=t_y$, which indicates that the system is neither localized nor extended, but critical. In comparison with the quantity $z$, the scaling of $\sigma_{yy} \Gamma$ in the limit $T=0$ and small $\Gamma$ regime can provide a clearer indicator of different regimes. For $t_y<t_x$, the conductance $\sigma_{yy} \Gamma$ decays exponentially with respect to $q$, implying localization in the $y$ direction as $q\rightarrow\infty$, as shown in Fig.~\ref{GS7M}(b). For $t_y>t_x$, the conductance $\sigma_{yy}\Gamma$ is independent of $q$, implying metallic behavior in this direction.  For $t_x=t_y$, the conductance exhibits a power-law decay with $q$, implying a critical regime.

As a comparison, we also plot $\langle C_2 \rangle$ and $\Delta v$ [Fig.~\ref{GS7M}(c,d)], discussed in detail in part A, on a similar scale. For these two quantities, the critical point $t_x=t_y$ shows a power-law decay as a function of $q$, in contrast with an exponential decay as a function of $q$ for the localized regime or the independence of $q$ in the extended regime, indicating that, at the quasiperiodic limit, the bulk transport is qualitatively stronger than that in the localized regime, but qualitatively weaker than that in the extended regime, justifying that the system is indeed critical at the isotropic point $t_x=t_y$.

\begin{figure}
\centerline{\includegraphics{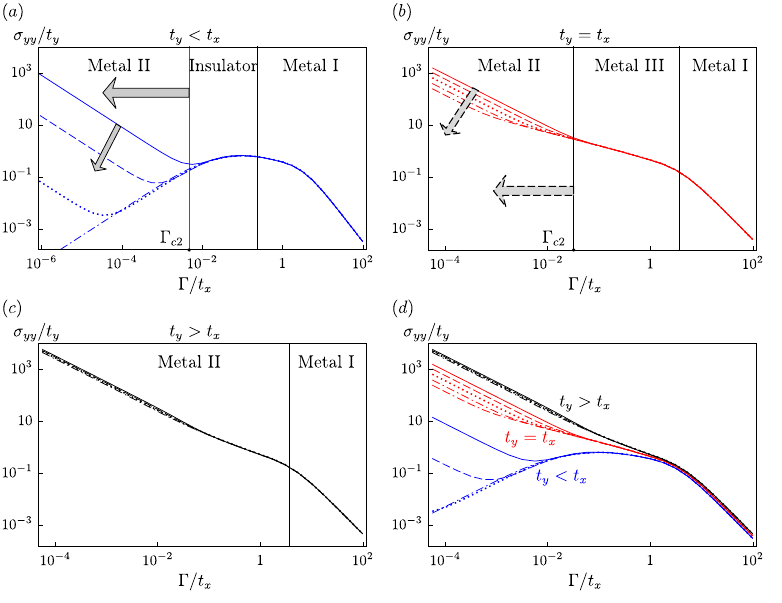}}
\caption{$\sigma_{yy}$ vs $\Gamma$, where $t_x=1$ and $t_y=1.2$ (a); $t_y=1$ (b) or $t_y=0.8$ (c), respectively. $\phi = 8/13, 13/21$, $21/34$, $34/55$ and $55/89$, corresponding to the solid, dashed, dotted, dashed-dotted and dashed-dashed-dotted lines, respectively. Blue, red and black colors correspond to three different parameter conditions $t_x>t_y$, $t_x=t_y$ and $t_x<t_y$, respectively. Gray arrows indicate the behavior as $q$ increases. Two metallic regions (Metal I/II) with distinct origins are shown in (a,b,c). The middle region denotes an insulator phase in $y$ direction for $t_x<t_y$ (a), and a critical metal phase for $t_x=t_y$ (b). $E_F$ is set within a band. (d) The combined plot of (a-c) for a comparison.}
\label{GS8M}
\end{figure}

\subsection{The isotropic case with $t_x=t_y$}
In this section, we further establish the essential differences between the isotropic case $t_x = t_y$ and the anisotropic case by investigating the conductivity $\sigma_{yy}$ at $T=0$ with finite relaxation rate $\Gamma$, which reveals intriguing effects. We begin by focusing on the aforementioned asymptotic behavior ($T=0$, small $\Gamma$) of the conductivity, i.e., $\sigma_{yy}$ decays exponentially, decays algebraically or remains unchange as $q$ increases at $t_x>t_y$, $t_x=t_y$, or $t_y>t_x$, respectively. This behavior can be inspected both in Fig.~\ref{GS7M}(b) and Figs.~\ref{GS8M}(a-c). As shown in Figs.~\ref{GS8M}(a,b), the gray tilted arrows indicate the direction of increasing $q$, which is set to be $q=13, 21, 34, 55$ and $89$. In the anisotropic regime with $t_x>t_y$ [Fig.~\ref{GS8M}(a)], the spacing between the straight lines in the Metal II region increases exponentially, indicating an exponential scaling of conductivity versus $q$. In contrast, at the isotropic regime with $t_x=t_y$ [Fig.~\ref{GS8M}(b)], the lines are equally spaced, indicating a power-law scaling of conductivity versus $q$ in this region. Lines in Fig.~\ref{GS8M}(c) overlap, indicating that the conductivity is independent of $q$. Beyond the small $\Gamma$ limit, we found that the critical relaxation rate $\Gamma_{c2}$, which denotes the leftmost phase boundary in Fig.~\ref{GS8M}(a) and Fig.~\ref{GS8M}(b), decreases exponentially and algebraically, respectively. Most nontrivially, in the middle region with $\Gamma_{c2}<\Gamma<\Gamma_{c1}$, the original insulator for $t_x>t_y$ turns into a critical metal (Metal III) at the isotropic point $t_x=t_y$ under the interplay between asymptotic quasiperiodicity and relaxation. In the critical metal phase, the conductivity satisfies $\sigma \sim \Gamma^{-0.5}$, which is a unique behavior not seen in the localized or extended phases. Similar to the above discussion, these results suggest that at the isotropic point $t_x=t_y$, the system is deeply connected to the critical phase in the quasiperiodic limit. For comparison, we also plot a graph simultaneously showing the three cases ($t_x>t_y$, $t_x=t_y$ and $t_x<t_y$) together [Fig.~\ref{GS8M}(d)].